\documentclass[aps,pra,twocolumn,10pt,showpacs]{revtex4-1}

\usepackage{amsmath}
\usepackage{graphicx}
\usepackage{times}
\usepackage{amssymb}
\usepackage{hyperref}
\usepackage{textcomp}
\usepackage{xcolor}
\usepackage{pdfpages}

\begin{document}
\title{Optimal Squeezing in the Resonance Fluorescence of Single Photon Emitters}
\author{P. Gr\"unwald}
\email{Electronic address: peter.gruenwald2@uni-rostock.de}
\author{W. Vogel}
\affiliation{Arbeitsgruppe Theoretische Quantenoptik, Institut f\"ur Physik, Universit\"at Rostock, D-18055 Rostock, Germany}
\date{\today}

\begin{abstract}
The possibilities and perspectives of squeezed light emission are studied for coherently driven single photon sources, such as atoms and quantum dots.
Maximal squeezing is realized, if the electronic subsystem of the emitter is in a pure quantum state. The purification is achieved by 
using a cavity as a second decay channel, besides the incoherent coupling to the electromagnetic vacuum. For realistic cavities this yields a purity of the electronic state of more than 99\%. Aside from numerical calculations, we also derive approximate analytical results. Based on the approximations, effects are studied which originate from the environment of the emitter, including radiationless dephasing and incoherent pumping of the emitter and the cavity mode. The fragility of squeezing against decoherence is substantially reduced, so that squeezing persists even under hostile conditions. The measurement of squeezing from such light sources is also considered.
\end{abstract}


\maketitle

\section{Introduction.}
Light fields having a cutoff in the photon statistics are known to exhibit nonclassical features. 
Of particular interest are single photon emitters (SPEs), whose emitted field consists of one photon in a properly defined mode volume. A single, laser-driven two-level atom was the first SPE under study~\cite{carm,kimble}. The first experimental demonstration of the  
quantum nature of light through the photon antibunching effect was realized with a low-density atomic beam~\cite{ki-dag-ma}. In a related experiment a sub-Poisson photon statistics was observed~\cite{short}.
Later on, photon antibunching has also been demonstrated with single trapped ions~\cite{walther,toschek}.
Nowadays, artificial atoms, such as quantum dots in semiconductor micro-systems, are established SPEs, showing photon antibunching and sub-Poisson photon statistics~\cite{MichlerExp,MichlerBook,Shih-1,Shih-2}.

Squeezed light was also predicted to occur in the resonance fluorescence of driven SPEs~\cite{walls}. Squeezing can also be realized and enhanced in the fluorescence of many atoms in different scenarios. They include the regular arrangement of the atoms~\cite{vowe}; the detection in the forward direction with respect to the pump-beam~\cite{heire}; and the bistability in a strong driving field~\cite{Reid}. The latter two cases could be experimentally demonstrated~\cite{LuBali,Raizen}. Recently it was shown, that the output field of a driven cavity, containing an atom, shows weak squeezing~\cite{Rempe}.

A direct observation of squeezing from a single driven SPE could neither be demonstrated in atoms nor in quantum dots; hence, this is an issue of fundamental interest. The standard method for detecting  squeezing is balanced homodyne detection. In this case, the observable effects in the resonance fluorescence are tiny due to the small collection efficiency of the light field.
Based on homodyne correlation measurements, efficient measurement techniques were proposed~\cite{vogel,vogel95,shchukin}, which are not limited by the small collection efficiency. The feasibility of such techniques has been demonstrated recently, in resonance fluorescence of a single trapped ion~\cite{Blatt}.

Recently we have shown that squeezed light in the fluorescence of an atom can be optimized~\cite{Sq-Gruen}. The maximal possible squeezing depends on the excitation of the emitter. It is achieved when the  electronic subsystem of the emitter is in a pure quantum state. Specifically, we found, that the maximal attainable squeezing is twice as large as for an atom in free space. This indicates the unfavorable situation of an atom which only couples to the vacuum field modes. We presented a purification scheme, solely based on a second coupling of the atom to a cavity field, which yields over 99\% purity and 94\% of the maximal possible squeezing. The fluorescence light must be observed out the side of the cavity, so that the latter acts as a passive environment. We predicted the squeezing to be substantially less fragile against radiationless dephasing.
 
In the present paper we study the fluorescence of general SPEs and investigate the possible squeezing and its limitations.
We derive an analytical approximation of the purification procedure, which allows us to interpret the underlying physics. Furthermore, based on the approximation we analyze different environmental effects, as they occur in a quantum dot in a semiconductor microcavity. The results indicate, that squeezing  persists even under certain perturbations, so that quantum dots in semiconductor microcavities may be a promising source of squeezed light, to be used in integrated optical systems.
Finally, we deal with a simple method to detect the emitted squeezed light. 

The article is organized as follows. In Sec.~\ref{sec.II} we study the squeezing in the resonance fluorescence of a SPE. The purification method is considered in Sec.~\ref{sec.III}. In Sec.~\ref{sec.IV} we introduce our analytical approximations, which we apply in Sec.~\ref{sec.V} to the environmental effects of dephasing and incoherent pumping. The possibility to detect the squeezing is analyzed in Sec.~\ref{sec.VI}. Finally, in Sec.~\ref{sec.VII} we give some conclusions and an outlook.

\section{Fluorescence Squeezing of SPEs}\label{sec.II}
A light field $\hat E$ (dependencies on space and time are suppressed throughout the paper, unless needed) is squeezed, if its variance $\langle(\Delta\hat E)^2\rangle$ is below the variance in the vacuum state, $\langle(\Delta\hat E)^2\rangle_\text{vac}$. Equivalently, the normally ordered variance,
\begin{equation}
  \langle:(\Delta\hat E)^2:\rangle=\langle(\Delta\hat E)^2\rangle-\langle(\Delta\hat E)^2\rangle_\text{vac},\label{eq.DelEgen}
\end{equation}
attains negative values, $\langle:(\Delta\hat E)^2:\rangle < 0$.
The ``$:\cdots:$'' prescription denotes normal ordering. 
The advantage of the normal ordering prescription is, that it separates free fields, $\hat E_\text f$, from source fields, $\hat E_\text s$, where $\hat E=\hat E_\text f+\hat E_\text s$. If only the atomic source field hits the detector, the free field does not contribute to normally ordered correlation functions, so that
\begin{equation}
  \langle:(\Delta\hat E)^2:\rangle=\langle:(\Delta\hat E_\text s)^2:\rangle.
\end{equation}
A detailed treatment of the source fields, in particular for the case of atomic resonance fluorescence, can be found in~\cite{WelVo}.
In the following we will omit the source field index 's'. 

Based on the dipole- and rotating-wave approximation for the light matter coupling, a quasi-monochromatic source field can be written as an effective single-mode field,
\begin{equation}
  \hat E=\chi(\hat b e^{-i\varphi}+\hat b^\dagger e^{i\varphi}),
\end{equation}
$\hat b$ ($\hat b^\dagger$) being the annihilation (creation) operator of the matter excitation inducing the dipole, while $\varphi$ describes the phase of the field. The scaling factor $\chi$ has the dimension of the electric field strength.
For this case, the normally ordered field variance becomes
\begin{equation}
  \begin{split}
    \langle:(\Delta\hat E)^2:\rangle=&2|\chi|^2\big[\langle\hat b^\dagger\hat b\rangle-|\langle\hat b\rangle|^2\\
				     &+\Re\{\langle\hat b^2\rangle-\langle\hat b\rangle^2)e^{-2i\varphi}\}\big].
  \end{split}
\end{equation}
Optimizing with respect to $\varphi$, we obtain the phase $\varphi_+\,(\varphi_-)$ of maximal~(minimal) normally ordered variance, $\langle:(\Delta\hat E)^2:\rangle_{\pm}$, which read as
\begin{align}
  e^{-2i\varphi_\pm}&=\pm \sqrt{\frac{\langle\hat b^2\rangle-\langle\hat b\rangle^2}{\langle\hat b^{\dagger2}\rangle-\langle\hat b^\dagger\rangle^2}}\,,\\
  \langle:(\Delta\hat E)^2:\rangle_{\pm}&=2|\chi|^2\Big[\langle\hat b^\dagger\hat b\rangle-\big|\langle\hat b\rangle|^2\pm|\langle\hat b^2\rangle-\langle\hat b\rangle^2\big|\Big].\label{eq.SMvar}
\end{align}
In the following, except Sec.\ref{sec.VI}, we only consider the phase optimized, minimal field fluctuation, which is the maximal squeezing. Hence, we omit the index '$-$'.

Let our SPE be a general two-level system in an arbitrary environment. In such a scenario, the source fields are proportional to the atomic operators, see~\cite{WelVo}. We can thus identify the annihilation and creation operators with the atomic flip operators $\hat A_{ij}=|i\rangle\langle j|$ ($i,j=1,2$), with $|1\rangle$ and $|2\rangle$
being the ground and excited state, respectively. The source field reads as
\begin{equation}
  \hat E=\chi(\hat A_{12}e^{-i\varphi}+\hat A_{21} e^{i\varphi}),
\end{equation}
and the minimal normally ordered variance can be written as
\begin{align}
  \langle:(\Delta\hat E)^2:\rangle=&2|\chi|^2(\langle\hat A_{22}\rangle-2|\langle\hat A_{12}\rangle|^2)\label{eq.Atvarsol1}.
\end{align}
For any atomic excitation, $\langle\hat A_{22}\rangle$, maximal squeezing is obtained for maximal atomic coherence, $|\langle\hat A_{12}\rangle|^2$.  

The expectation values in Eq.~(\ref{eq.Atvarsol1}) are readily derived from the  density operator of the SPE,
$\hat \sigma = \sum_{ij} \sigma_{ij} \hat A_{ij}$. The density matrix $\sigma$ reads as
\begin{equation}
  \sigma=\left(\begin{array}{cc}
		    \langle\hat A_{11}\rangle & \langle\hat A_{21}\rangle\\ \langle\hat A_{12}\rangle & \langle\hat A_{22}\rangle
                   \end{array}
\right).
\end{equation}
Combining the positive semi-definiteness of quantum states, $\det \sigma \ge 0$, with the completeness relation of the SPE, we get
\begin{equation}
  |\langle\hat A_{12}\rangle|^2\leq\langle\hat A_{11}\rangle\langle\hat A_{22}\rangle=\langle\hat A_{22}\rangle-\langle\hat A_{22}\rangle^2,\label{eq.CSI}
\end{equation} 
which yields the maximal coherence, $|\langle\hat A_{12}\rangle|^2$, as function of the excitation $\langle\hat A_{22}\rangle$.

Using this result in Eq.~(\ref{eq.Atvarsol1}), the minimal variance follows as
\begin{equation}
  \frac{\langle:(\Delta\hat E)^2:\rangle_{\text{min}}}{|\chi|^2}=2\langle\hat A_{22}\rangle(2\langle\hat A_{22}\rangle-1),\label{var-min}
\end{equation} 
 with the absolute minimum being given for $\langle\hat A_{22}\rangle=1/4$,
\begin{equation}
  \frac{\langle:(\Delta\hat E)^2:\rangle_{\text{abs}}}{|\chi|^2}=-\frac{1}{4}. \label{abs}
\end{equation}
The purity of the state of the SPE reads as
\begin{equation}
  \text{Tr}\{\hat \sigma^2\}=1-2(\langle\hat A_{22}\rangle-\langle\hat A_{22}\rangle^2-|\langle\hat A_{12}\rangle|^2).\label{eq.purity}
\end{equation}
Compared with Eq.~(\ref{eq.CSI}), a pure SPE state is equivalent to maximal SPE coherence and hence optimal squeezing. 

Knowing the maximal possible squeezing, the question appears: how much squeezing can one achieve in the fluorescence of a SPE in free space? Squeezing of the fluorescence light was predicted~\cite{walls,WelVo}, but not yet experimentally confirmed. Let $\omega_\text x$ be the frequency difference between ground and excited state. The pump laser field is in a coherent state of frequency $\omega_\text L=\omega_\text x-\delta_\text x$. The coupling strength between SPE and laser field is given by the Rabi-frequency $\Omega_\text R$, the laser phase is included in the phase of the source field. The SPE couples to the  vacuum modes (spontaneous emission) with rate $\Gamma$. In the frame rotating with $\omega_\text L$, the Hamiltonian and the master equation read as
\begin{align}
  \hat H_0&=\hbar\delta_\text x\hat A_{22}+\hbar\Omega_\text R(\hat A_{12}+\hat A_{21}),\label{eq.Ham0}\\
  \dot{\hat \sigma}&=\frac{1}{i\hbar}[\hat H_0,\hat\sigma]+\frac{\Gamma}{2}\mathcal{L}_{\hat A_{12}}[\hat\sigma],\label{eq.Mas0}\\
  \mathcal{L}_{\hat O}[\hat\sigma]&=2\hat O\hat\sigma\hat O^\dagger-\{\hat O^\dagger\hat O,\hat\sigma\}.\label{eq.Lind-gen}
\end{align}
This system can be solved analytically. The structure of the solution will be helpful for the discussion of environmental effects in the following sections. 

The steady state values of both excitation and coherence can be given by a single variable $z$, 
\begin{align}
  z&=\frac{\Omega_\text R^2}{(\tfrac{\Gamma}{2})^2+\delta_\text x^2},\label{eq.fs-z}\\
  \langle\hat A_{22}\rangle&=\frac{z}{1+2z},\quad|\langle\hat A_{12}\rangle|^2=\frac{z}{(1+2z)^2}.\label{eq.fscor}
\end{align}
Inserting these results into Eq.~(\ref{eq.purity}), we get
\begin{align}
  \text{Tr}\{\hat \sigma^2\}=1-2\langle\hat A_{22}\rangle^2.\label{eq.RF-pur}
\end{align}
The purity of the (stationary laser-driven) atom requires $\langle\hat A_{22}\rangle=0$, i.e. the SPE being in the ground state. In this case it obviously cannot emit fluorescent light.
With increasing excitation the purity of the SPE state diminishes. For saturation,  $\langle\hat A_{22}\rangle=1/2$ for $z\rightarrow\infty$, the state of the emitter is fully mixed, without any coherence. 
Thus, the regime of maximal squeezing cannot be reached in free space fluorescence. Squeezing can only be observed for low atomic excitation. The free space normally ordered variance, Eq.~(\ref{eq.Atvarsol1}), reads as
\begin{equation}
   \frac{\langle:(\Delta\hat E)^2:\rangle_{\text{fs}}}{|\chi|^2}=\frac{2z(2z-1)}{(1+2z)^2}.\label{eq.Varfs}
\end{equation}
Squeezed light is obtained for $z\leq1/2$, maximal squeezing in free space is realized for the parameters
\begin{align}
  z=\frac{1}{6},\ \text{or}\ \langle\hat A_{22}\rangle&=\frac{1}{8},\\
  \frac{\langle:(\Delta\hat E)^2:\rangle_{\text{fs,min}}}{|\chi|^2}&=-\frac{1}{8}\label{eq.SqfS}.
\end{align}

Altogether, in free-space fluorescence from a SPE the possible squeezing is limited by the impurity of the quantum state of the emitter. This impurity results from the coupling to the vacuum modes described by the master equation~(\ref{eq.Mas0}). For an optimization of squeezing in resonance fluorescence from a SPE, the task is to realize a purification of the atomic state for non-vanishing excitation. This can be achieved by a proper environment of the SPE, such as a cavity~\cite{Sq-Gruen}.
In the following we will provide analytical approximations, which will help to better understand such scenarios, with the aim to optimize the resistance of squeezing against various perturbations.

\section{Cavity induced Purification}\label{sec.III}
\begin{figure}[h]
  \includegraphics[width=7cm]{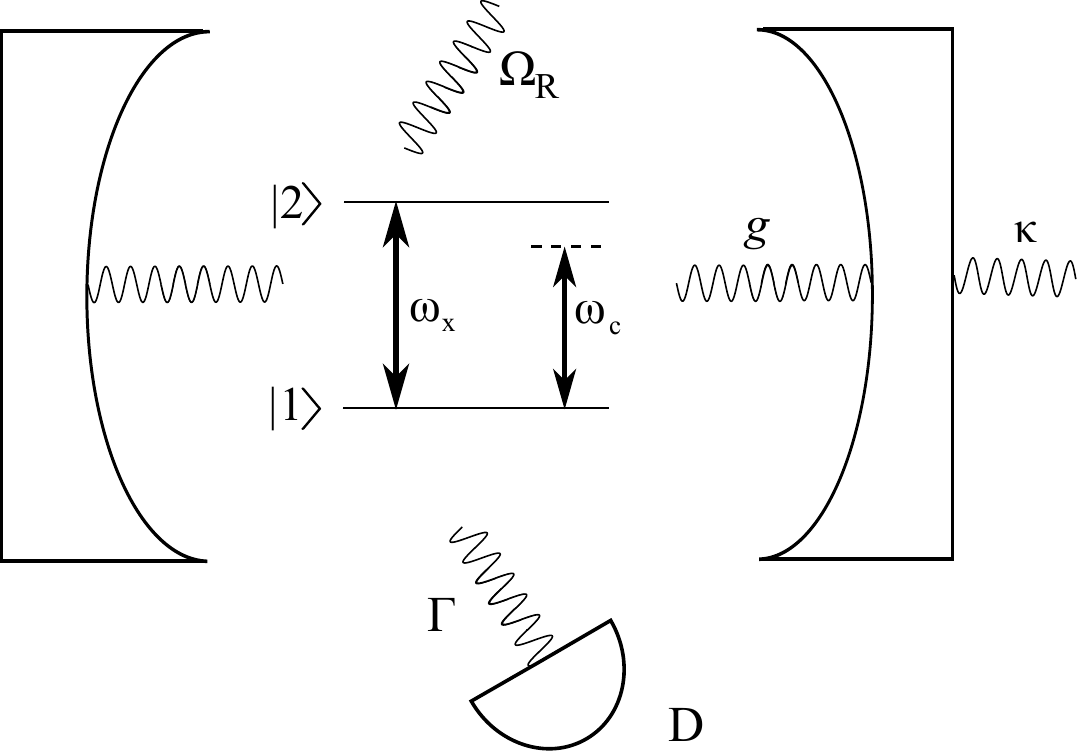}
  \caption{Sketch of the coherently driven SPE inside a lossy cavity. The fluorescent light is detected (D) out the side of the cavity. Wavy lines indicate light fields driving the SPE, emitted into the cavity, out of the cavity, or out the side of the cavity. Straight arrows indicate the frequencies of the SPE transition and the cavity mode.}\label{fig.sys}
\end{figure}

Pure states of SPEs have attracted significant attention over recent years due to their application as qubits in quantum information theory. Hence, protocols for purification~\cite{Bowles,Kiess}, and the determination of purity~\cite{Filip,Ekert,Nakaz} have been established. 
Recently we have shown that an optical cavity may act as a passive environment to purify the electronic state and to optimize squeezing of an atom undergoing resonance fluorescence~\cite{Sq-Gruen}. A sketch of our setup is given in Fig.~\ref{fig.sys}.

In addition to the system in Eqs.~(\ref{eq.Ham0})--(\ref{eq.Lind-gen}), the SPE is coupled to a single-mode cavity of frequency $\omega_\text c$, with coupling strength $g$. The cavity excitation is described by bosonic creation and annihilation operators, $\hat a^\dagger$ and $\hat a$, respectively, and the cavity has an emission rate $\kappa$. The full Hamiltonian for this system reads as
\begin{equation}
  \hat H=\hat H_0+\hbar\delta_\text c\hat a^\dagger\hat a+\hbar g(\hat a^\dagger\hat A_{12}+\hat A_{21}\hat a),\label{eq.Hamilton}
\end{equation}
where $\delta_\text c=\omega_\text c-\omega_\text L$. The density operator $\hat\varrho$ of the full system obeys the master equation
\begin{align}\label{eqmo}
    \frac{d\hat\varrho}{dt}&=\frac{1}{i\hbar}[\hat H,\hat\varrho]+\frac{\Gamma}{2}\mathcal{L}_{\hat A_{12}}[\hat\varrho]+\frac{\kappa}{2}\mathcal{L}_{\hat a}[\hat\varrho].
\end{align}
A general analytical solution of this system is unknown. For our numerical approach, we refer to appendix~\ref{sec.appnum}.

In the scenario discussed in~\cite{Sq-Gruen}, the SPE undergoes strong offresonant pumping, such that $\Gamma\ll \Omega_\text R,|\delta_\text x|$ and hence $z\approx\Omega_\text R^2/\delta_\text x^2$. As the cavity detuning is $|\delta_\text c|>|\delta_\text x|$ and large compared to the SPE-cavity coupling $g$, the intracavity-field is nearly unexcited, $\langle\hat a^\dagger\hat a\rangle\ll1$. However, following the argumentation in~\cite{FrQu,QuFr}, the intracavity field is enhanced if a fluorescence sideband hits the cavity frequency,
\begin{equation}
  \delta_\text c^2=(2\Omega_\text R)^2+\delta_\text x^2.\label{eq.cavres}
\end{equation}
In the following, the cavity frequency is tuned to the lower Rabi sideband of the SPE. For simplicity, we will denote this scenario as cavity resonance.

At such a cavity resonance, excitation and emission of the cavity are enhanced. The excitation of the cavity increases slightly, which is consistent with the argumentation in~\cite{FrQu}.  The cavity emission, however, increases substantially, due to $\kappa\gg\Gamma$. A similar situation was considered in~\cite{Carm11}, where steady-state inversion of a two-level atom in a cavity was predicted. In our scenario the cavity mode diverts a significant portion of the energy from the SPE, which would otherwise contribute to the fluorescent light. This yields a reduction of $\langle\hat A_{22}\rangle$. It is noteworthy, that for our system, a not too good cavity is preferable, in order to avoid a too strong backaction onto the SPE. On the other hand, the coupling of the cavity and SPE has to be strong enough to preserve the coherence of the SPE. 
The coherent part of the SPE, $|\langle A_{12}\rangle|^2$, is increased by the coupling to the cavity. Due to the decrease of $\langle\hat A_{22}\rangle$, from Eqs.~(\ref{eq.CSI}),~(\ref{eq.purity}) the SPE state is expected to be purified. 

The strength of the purification will be discussed in the next section. A critical condition in this setup is the requirement of $\kappa\gg\Gamma$ and $g\approx\kappa$, so that the SPE-cavity coupling significantly exceeds the spontaneous emission, $g\gg\Gamma$. In experiments~\cite{Hood}, a rate of
$g/\Gamma\approx23$ was realized, which will be used throughout the paper.

\begin{figure}[h]
  \includegraphics[width=4cm]{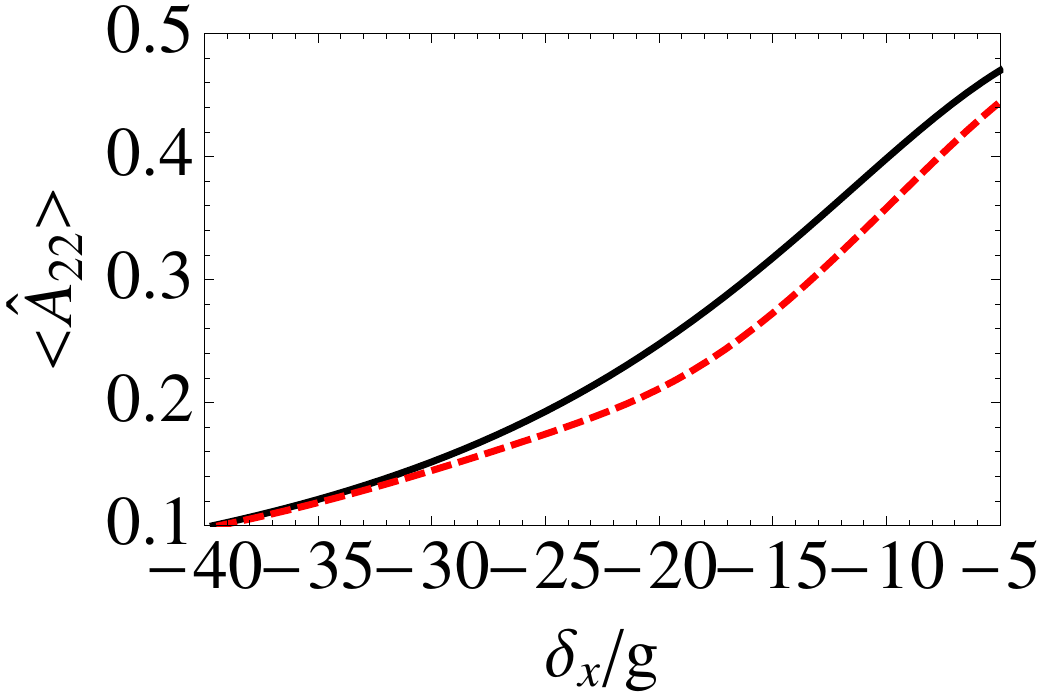}
\hspace{0.1cm}
  \includegraphics[width=4cm]{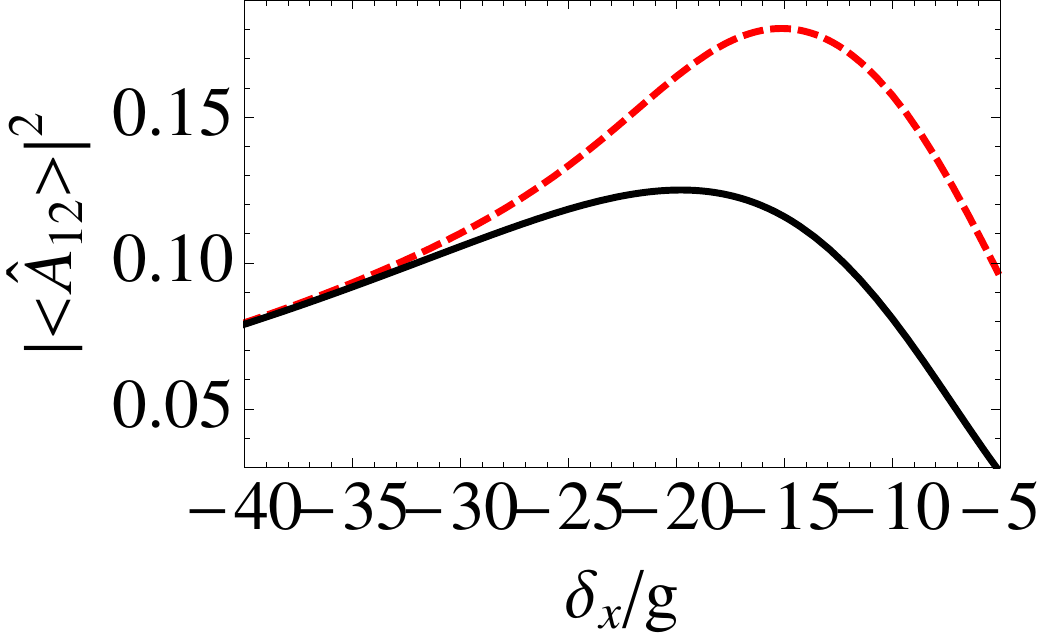}

\vspace{0.3cm}

  \includegraphics[width=4cm]{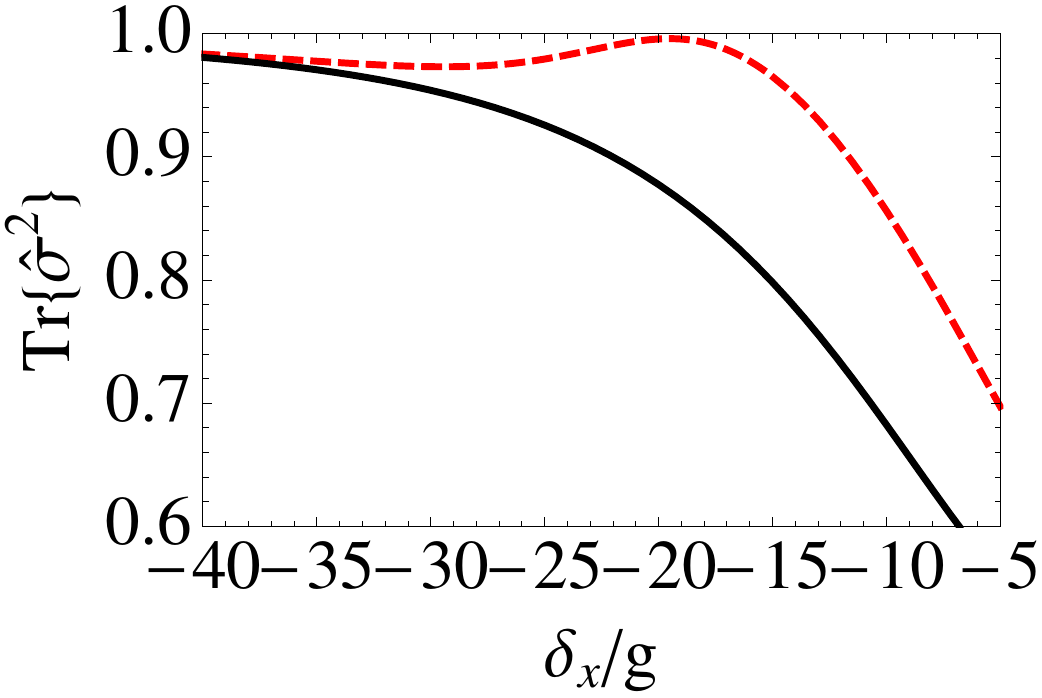}
\hspace{0.1cm}
  \includegraphics[width=4cm]{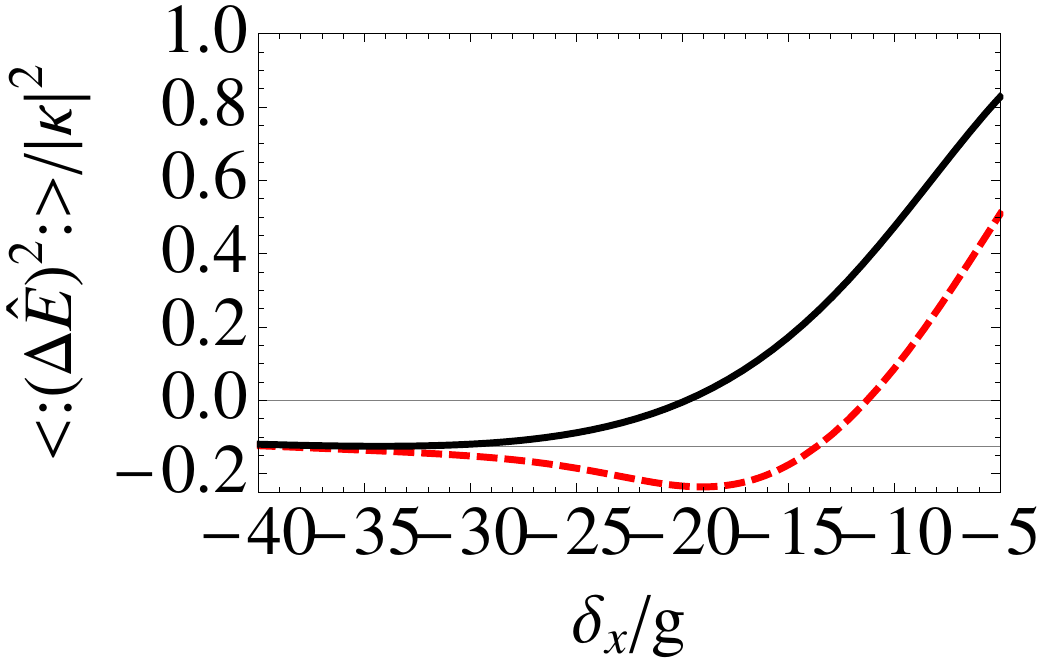}
  \caption{(color online). Comparison of the behavior of the SPE in free space (solid, black curve) and in the cavity (red, dashed curve). The excitation $\langle\hat A_{22}\rangle$ (left top), the   coherence $|\langle\hat A_{12}\rangle|^2$ (right top), the purity (left bottom) and the phase-optimized normally ordered field variance of the fluorescence (right bottom) are shown as a function of $\delta_\text x$. Two straight lines (right bottom) at $-1/8$ and $0$ indicate maximal and vanishing free-space squeezing, respectively. The parameters are: $\Omega_{\text R}/g=14$, $\kappa/g=1.58$, $\Gamma/g=1/23$, $\delta_\text c/g=-34$.}\label{fig.Sqmax}
\end{figure}

In~\cite{Sq-Gruen} we considered the following cavity scenario: $\Gamma=1/23g$,  $\Omega_\text R=14g$, $\delta_\text c=-34g$, and $\kappa=1.58g$. With these parameters, we numerically evaluate the systems parameters and the squeezing of the fluorescence in dependence on $\delta_\text x$, see Fig.~\ref{fig.Sqmax}. The sought cavity resonance is obtained for $\delta_\text x\approx-19g$, corresponding to $z\approx0.54$. Note that, for $z\geq1/2$ there is no squeezing in free space fluorescence.

At the cavity resonance the minimal normally ordered field variance is $-0.236$. This is more than 94\% of the maximum possible squeezing of $-1/4$. The purity Tr$\{\hat\sigma^2\}$ of the SPE subsystem even reaches a value of about 99.5\%. The SPE excitation of $\langle\hat A_{22}\rangle\approx0.220$ is reduced compared with its free space value, while the coherence drastically increases. 
Note that, for larger values of $g/\Gamma$ one could achieve even more than 99\% of the absolute squeezing limit, $\langle:(\Delta\hat E)^2:\rangle_{\text{abs}}$.

\section{Approximate analytical description}\label{sec.IV}
Now we will provide an analytical approximation to the system under study. Although its numerical precision is limited, it gives insight in the basic  physics of the cavity-assisted purification. Furthermore it will help to  predict environmental influences in the next section. All the predictions will be confirmed by numerical calculations.

In the steady state regime, we obtain from Eqs.~(\ref{eq.Hamilton}),~(\ref{eqmo})
the following exact relations:
\begin{align}
    \langle\hat a\rangle&=\frac{-ig}{i\delta_\text c+\tfrac{\kappa}{2}}\langle\hat A_{12}\rangle,\label{eq.ana-a}\\
    \left[i\delta_\text x{+}\tfrac{\Gamma}{2}\right]\langle\hat A_{12}\rangle&=ig(2\langle\hat A_{22}\hat a\rangle{-}\langle\hat a\rangle){-}i\Omega_\text R(1{-}2\langle\hat A_{22}\rangle),\label{eq.ana-A12}\\
    \langle\hat A_{22}\rangle&=\frac{2\Omega_\text R}{\Gamma}\Im\langle\hat A_{21}\rangle-\frac{\kappa}{\Gamma}\langle\hat a^\dagger\hat a\rangle.\label{eq.ana-A22}
\end{align}
The proportionality between the coherent amplitudes of SPE and intracavity field, cf. Eq.~(\ref{eq.ana-a}), is connected to the Purcell-effect~\cite{Purcell1,Carm-Book}. Combining Eq.~(\ref{eq.ana-a}) and (\ref{eq.ana-A12}), we obtain
\begin{align}
    V\langle\hat A_{12}\rangle&=-i\Omega_\text R(1-2\langle\hat A_{22}\rangle)+2ig\langle\hat A_{22}\hat a\rangle,\label{eq.VA12}\\
    V&=i\delta_\text x+\tfrac{\Gamma}{2}+\frac{g^2}{i\delta_\text c+\tfrac{\kappa}{2}}.
\end{align}
The term proportional to $g^2$ in $V$ is directly related to the Purcell-factor and enhances the total emission from the SPE. The term proportional to $\Omega_\text R$ in Eq.~(\ref{eq.VA12}) is exactly the same as in the free-space resonance fluorescence. The only higher order term is $\langle\hat A_{22}\hat a\rangle$. It contributes significantly when both the SPE and the intracavity field are excited. Due to the very weak cavity excitation in our scenario, this is the smallest contribution, and is neglected in the following. The coherence of the SPE reduces to
\begin{align}
    \langle\hat A_{12}\rangle&=\frac{-i\Omega_\text R}{V}(1-2\langle\hat A_{22}\rangle).
\end{align}
It resembles the free-space result, with a change of the scaling factor $V$. For the parameters of the above simulations, the change of $V$ 
in Fig.~\ref{fig.Sqmax} relative to free space, $g=0$, is negligible.

The term proportional to $\Omega_\text R$ in Eq.~(\ref{eq.ana-A22}) also resembles the free-space term. The second term, 
\begin{equation}
R=\frac{\kappa}{\Gamma}\langle\hat a^\dagger\hat a\rangle>0,\label{eq.def-R}
\end{equation}
describes the sharing of the excitation between SPE and cavity mode discussed in the previous section. While $\langle\hat a^\dagger\hat a\rangle\ll1$, it is scaled up by a factor of $\kappa/\Gamma\approx36$, making it a significant contribution. This quantity $R$ causes the purification. From now on, we will call $R$ the purification rate.

Considering $R$ as a parameter of the calculations, we end up with the following two equations
\begin{align}
    \langle\hat A_{22}\rangle&=\frac{2\Omega_\text R}{\Gamma}\Im\langle\hat A_{21}\rangle-R,\\
    \langle\hat A_{12}\rangle&=\frac{-i\Omega_\text R}{V}(1-2\langle\hat A_{22}\rangle).
\end{align}
Inserting these results into each other, we can conclude
\begin{align}
  \Im\langle\hat A_{21}\rangle&=\frac{\Omega_\text R}{|V|^2}\Re[V](1-2\langle\hat A_{22}\rangle),\\
  \langle\hat A_{22}\rangle&=\frac{2\Omega_\text R}{\Gamma|V|^2}\Re[V](1-2\langle\hat A_{22}\rangle)-R,\\
  \Rightarrow\ \langle\hat A_{22}\rangle&=\frac{\tilde z-R}{1+2\tilde z},\quad \tilde z= \frac{2\Omega_\text R^2}{\Gamma|V|^2}\Re[V],\\
  |\langle\hat A_{12}\rangle|^2&=\frac{\Gamma}{2\Re[V]}\frac{\tilde z(1+2R)^2}{(1+2\tilde z)^2}.
\end{align}
For $g\rightarrow0$, we get the free space case of $\Re[V]=\Gamma/2$ and $\tilde z=z=\Omega_\text R^2/|V|^2$. As stated above, 
$V$ does change marginally for our parameters. We can set $\Re[V]\approx\Gamma/2$ and $\tilde z\approx z$ to obtain
\begin{align}
	\langle\hat A_{22}\rangle&=\frac{z-R}{1+2z}<\frac{z}{1+2z},\label{eq.ana-A22sol}\\
  |\langle\hat A_{12}\rangle|^2&=\frac{z(1+2R)^2}{(1+2z)^2}>\frac{z}{(1+2z)^2}.\label{eq.ana-A12sol}
\end{align}
The positivity of $R$ diminishes the excitation of the SPE at the expense of  increasing its coherence. Even a small cavity excitation $\langle\hat a^\dagger\hat a\rangle$, scaled up by the prefactor in $R$, yields a substantial purification of the quantum state of the SPE. 

Inserting these approximations into Eq.~(\ref{eq.Atvarsol1}), we obtain 
in the cavity-assisted setup the result for the minimal field fluctuation, that is for maximal squeezing:
\begin{align}
  \frac{\langle:(\Delta\hat E)^2:\rangle_\text{cav}}{|\chi|^2}=&2(\langle\hat A_{22}\rangle-2|\langle\hat A_{12}\rangle|^2)\\
  =&\frac{\langle:(\Delta\hat E)^2:\rangle_\text{fs}}{|\chi|^2}-\frac{2R}{1+2z}\left(1+\frac{8(1+R)}{1+2z}\right).\label{eq.Sq-ana}
\end{align}
Here we have used the expression $\langle:(\Delta\hat E)^2:\rangle_\text{fs}$  according to Eq.~(\ref{eq.Varfs}). As expected, we have a clear decrease of the normally ordered variance, or an increase of squeezing, as the second term, proportional to $R$, is always positive.  

\section{Environmental Disturbances}\label{sec.V}

Based on the above approximations, let us study the important problem of environmental disturbances. 
We will consider three types of disturbances. Nonradiative or pure dephasing is caused by laser fluctuations or by atomic motion. In semiconductor microcavities containing quantum dots, two types of incoherent gains exist, either for the quantum dot or the cavity field~\cite{Khitrova,Finley,Michler}. They are caused by the interaction of the quantum dot with phonons, which will be modeled by Lindblad terms. Alternative descriptions of phonon-induced dephasing are given, e.g., in~\cite{Maju}. Semiconductor microcavities are currently immensely studied. They can be useful as nonclassical light sources in integrated optical systems. Our results will indicate that squeezing persists even under strong environmental disturbances.

\subsection{Nonradiative dephasing}
Nonradiative dephasing or pure dephasing is the radiationless decay of coherence of a system. In case of a SPE this obviously destroys the squeezing. Let us first reconsider the impact of pure dephasing in free space~\cite{WelVo}, and then compare it with the cavity-assisted squeezing scenario.

In addition to dephasing due to radiative damping, let there be radiationless dephasing described by the rate $\Gamma_\text D$.
We supplement the equations of motion~(\ref{eq.Mas0}) for the atom in free space by another Lindblad-term,
 \begin{equation}
      \frac{d\hat\rho}{dt}=\frac{1}{i\hbar}[\hat H_0,\hat\rho]+\frac{\Gamma}{2}\mathcal{L}_{\hat A_{12}}[\hat\rho]+\frac{\Gamma_\text D}{2}\mathcal{L}_{\hat A_{22}}[\hat \rho].
 \end{equation}
The additional dephasing only enhances the decay of the off-diagonal matrix elements of the density operator, that is, of the coherence of the SPE. We can again solve this system analytically and obtain
\begin{align}
  z_\text D&=(1+\tfrac{\Gamma_\text D}{\Gamma})\frac{\Omega_\text R^2}{(\tfrac{\Gamma+\Gamma_\text D}{2})^2+\delta_\text x^2},\\
  \langle\hat A_{22}\rangle&=\frac{z_\text D}{1+2z_\text D},\\
  |\langle\hat A_{12}\rangle|^2&=\frac{1}{1+\tfrac{\Gamma_\text D}{\Gamma}}\frac{z_\text D}{(1+2z_\text D)^2}.
\end{align}
Structurally, the solution for $\langle\hat A_{22}\rangle$ resembles the case without pure dephasing, with a scaled value $z_\text D$. In our scenario of large detuning, the variation of the denominator in $z_\text D$ is negligible, 
\begin{equation}
  z_\text D\approx(1+\tfrac{\Gamma_\text D}{\Gamma})\frac{\Omega_\text R^2}{\delta_\text x^2}=(1+\tfrac{\Gamma_\text D}{\Gamma})z.\label{eq.zd}
\end{equation}
With increasing dephasing rate $\Gamma_\text D$ the atomic excitation increases, while the coherence decreases as
\begin{equation}
  |\langle\hat A_{12}\rangle|^2\approx\frac{z}{(1+2z_\text D)^2}\label{eq.fsA22deph}.
\end{equation}
As the excitation increases, the pumping $\Omega_\text R$ has to be reduced to preserve squeezing. For $\Gamma_\text D=\Gamma$ squeezing vanishes as
\begin{equation}
 \frac{\langle:(\Delta\hat E)^2:\rangle_\text{fs}}{|\chi|^2}=\left(\frac{2z_\text D}{1+2z_\text D}\right)^2>0.\label{eq.fs-dep-sq}
\end{equation}
This limit for squeezing can be physically understood, as the time needed to emit a photon is as long as the coherence time of the emitted light.

In the cavity assisted fluorescence scenario, we repeat the calculations from Eqs.~(\ref{eq.ana-a})-(\ref{eq.ana-A12sol}), and obtain the following differences. The parameters $V$ and $z$ are changed to 
\begin{equation}
	V_\text D=V+\frac{\Gamma_\text D}{2},\quad z_\text D= \frac{2\Omega_\text R^2}{\Gamma|V_\text D|^2}\Re[V_\text D].\label{eq.dep-par}
\end{equation}
Again the real and imaginary part of $V_\text D$ are only marginally different from the free space values with pure dephasing. The purification rate $R$ on the other hand remains unchanged, as neither the excitation of the SPE nor the cavity are directly coupled to $\Gamma_\text D$.
The SPE averages then read as
\begin{align}
  \langle\hat A_{22}\rangle&=\frac{z_\text D-R}{1+2z_\text D},\label{eq.anadep-A22sol}\\
  |\langle\hat A_{12}\rangle|^2&=\frac{1}{1+\tfrac{\Gamma_\text D}{\Gamma}}\frac{z_\text D(1+2R)^2}{(1+2z_\text D)^2}.\label{eq.anadep-A12sol}
\end{align}
Similar to the case without radiationless dephasing, Eqs.~(\ref{eq.ana-A22sol}), and (\ref{eq.ana-A12sol}), the excitation $\langle\hat A_{22}\rangle$ is diminished while $|\langle\hat A_{12}\rangle|^2$ is enhanced by the positivity of $R$. Combining them to obtain the squeezing, Eq.~(\ref{eq.Atvarsol1}), we may compare with the result~(\ref{eq.Sq-ana}) for $\Gamma_\text D=0$, 
\begin{equation}
\begin{split}
  \frac{\langle:(\Delta\hat E)^2:\rangle_\text{cav}}{|\chi|^2}=&\frac{\langle:(\Delta\hat E)^2:\rangle_\text{fs}}{|\chi|^2}\\
  &-\frac{2R}{1+2z_\text D}\left(1+\frac{1}{1+\tfrac{\Gamma_\text D}{\Gamma}}\frac{8(1+R)}{1+2z_\text D}\right).\label{eq.Sqdep-ana}
\end{split}
\end{equation}
The second term in the brackets is now diminished by the dephasing prefactor.

Our results reveal that cavity-assisted purification increases the stability of squeezing against dephasing. 
The enhancement of the coherence is given by the ratio $(1+2R)^2$ to $1+\Gamma_\text D/\Gamma$ in Eq.~(\ref{eq.anadep-A12sol}). As $R\ll1$ (see above), one might not expect a significant effect. However, the dephasing also affects the intracavity excitation. The radiationless dephasing suppresses the coherence of the SPE, while it does not modify the coupling strength $g$ to the cavity. As the SPE is stronger excited for increasing $\Gamma_\text D$, $\langle\hat a^\dagger\hat a\rangle$ and hence $R\propto\langle\hat a^\dagger\hat a\rangle$ substantially increases at the cavity resonance. 

\begin{figure}[h]
  \includegraphics[width=8cm]{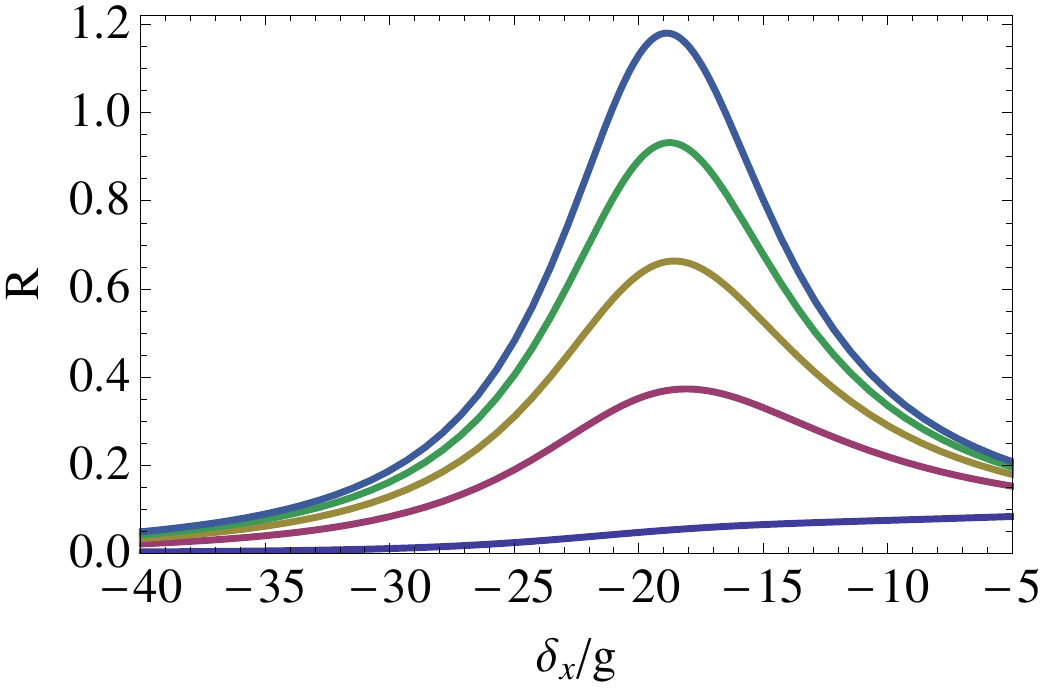}
  \caption{(color online). Purification rate $R$ over $\delta_x$ for different dephasing rates $\Gamma_\text D$. From bottom to top: $\Gamma_\text D/\Gamma=0,2,4,6,8$. All other parameters are as in Fig.~\ref{fig.Sqmax}.}\label{fig.adaDeph}
\end{figure}

The purification rate is depicted in Fig.~\ref{fig.adaDeph}. We see, that the increase of $R$ at the cavity resonance becomes more pronounced with increasing $\Gamma_\text D$. 

\begin{figure}[h]
  \includegraphics[width=8cm]{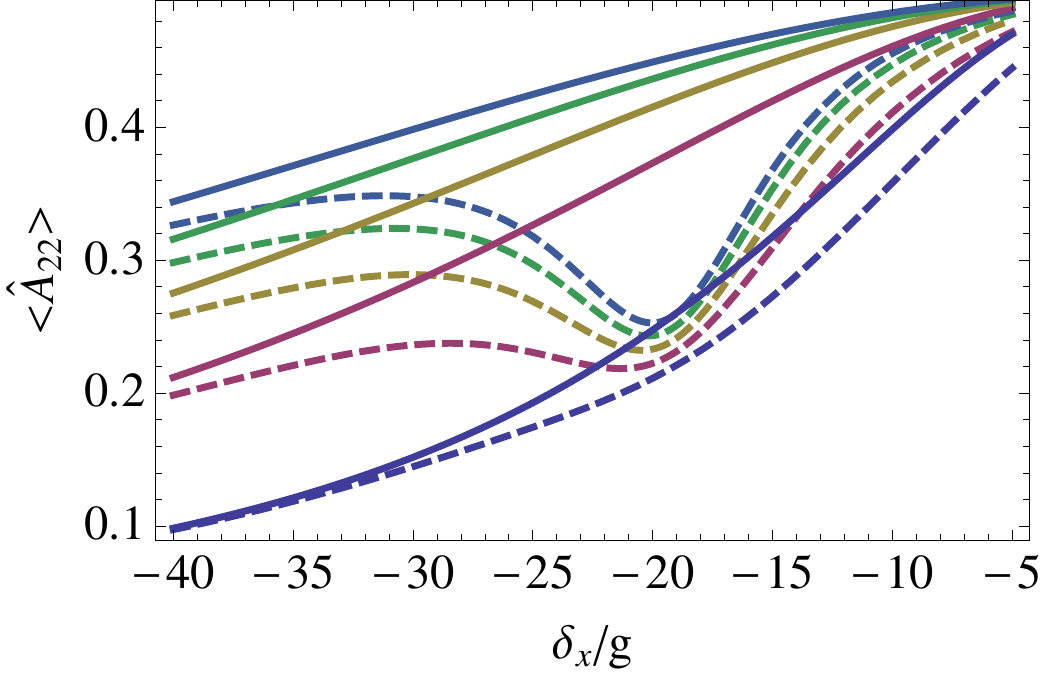}
  \caption{(color online). SPE excitation $\langle\hat A_{22}\rangle$ over $\delta_x$ for different dephasing rates $\Gamma_\text D$. The solid lines represent the free space case, the dashed ones the corresponding cavity assisted scenario. From bottom to top (for each solid and dashed lines separately): $\Gamma_\text D/\Gamma=0,2,4,6,8$. All other parameters are as in Fig.~\ref{fig.Sqmax}.}\label{fig.A22Deph}
\end{figure}

In Fig.~\ref{fig.A22Deph}, we compare the SPE excitation for different dephasing rates with and without cavity-assisted purification.
In the latter case, the excitation of the SPE is suppressed at the cavity resonance even below the free space value. 
The coherence near the resonance remains almost constant, as the terms $(1+2R)^2$ and $1+\Gamma_\text D/\Gamma$ in Eq.~(\ref{eq.anadep-A12sol}) are nearly equal.

\begin{figure}[h]
  \includegraphics[width=8cm]{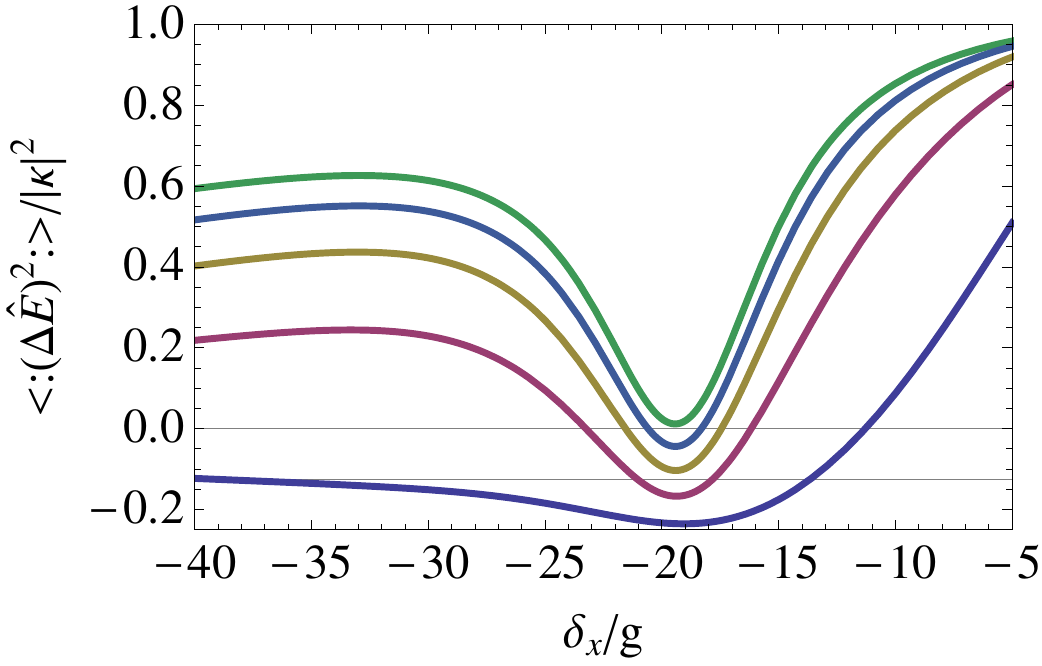}
  \caption{(color online). Squeezing of the SPE fluorescence over $\delta_\text x$ for different dephasing rates $\Gamma_\text D$. From bottom to top: $\Gamma_\text D/\Gamma=0,2,4,6,8$. All other parameters are as in Fig.~\ref{fig.Sqmax}. The horizontal lines indicate maximal free space squeezing (-1/8) and vanishing squeezing (0).}\label{fig.Deph}
\end{figure}

These effects imply, that the resistance of squeezing against dephasing is significantly enhanced. The phase-optimized normally ordered variance Eq.(\ref{eq.Atvarsol1}) for different dephasing in the cavity setup is shown in Fig~\ref{fig.Deph}. Due to the behavior of $\langle\hat A_{22}\rangle$ and $|\langle \hat A_{12}\rangle|^2$, the suppression of the field noise sensitively depends on $\delta_\text x$. 
For $\Gamma_\text D<3.24\Gamma$, the minimal variance is still below $-1/8$, being the maximal squeezing in free space.  The squeezing in the cavity setup under study vanishes for $\Gamma_\text D\approx7.47\Gamma$. 
This value is, however, not the actual limit. From Eqs.~(\ref{eq.fsA22deph}), (\ref{eq.anadep-A22sol}) and the increase of $R$, it seems reasonable to look for squeezing at lower pump rates $\Omega_\text R$. For lower pumping, the emitter frequency $\omega_{\text x}$ shifts towards the cavity frequency $\omega_{\text c}$, cf. Eq.~(\ref{eq.cavres}). While the squeezing in this region is not as strong as in Fig.~\ref{fig.Deph}, it is even more persistent. As an example, for $\Omega_\text R=g$, squeezing still persists for $\Gamma_\text D=19\Gamma$.

These findings are of great interest for condensed matter systems, where dephasing plays a significant role~\cite{Khitrova,Finley,Michler,Dousse}. The observation of  coherence effects, such as squeezing, under these hostile conditions is a demanding task. Note that, 
the needed variation of $\Omega_\text R$ and $\delta_\text x$ can be easily realized for a semiconductor quantum dot inside a cavity. 

\subsection{Incoherent Pumping of SPE}
The light emitted by a quantum dot in a semiconductor first passes the medium, where it excites phonons. The phonons can incoherently drive the quantum dot. The  incoherent pumping of the SPE will be included by a rate $P_\text x$. We will again start to consider the corresponding effects in free space, before analyzing the SPE  inside the cavity.

The free space master equation~(\ref{eq.Mas0}) is supplemented with another Lindblad term for the incoherent pumping,
\begin{equation}
      \frac{d\hat\rho}{dt}=\frac{1}{i\hbar}[\hat H_0,\hat\rho]+\frac{\Gamma}{2}\mathcal{L}_{\hat A_{12}}[\hat\rho]+\frac{P_\text x}{2}\mathcal{L}_{\hat A_{21}}[\hat \rho].
\end{equation}
The solutions in the steady state now read as
\begin{align}
  z_\text x&=\frac{\Omega_\text R^2}{(\tfrac{\Gamma+P_\text x}{2})^2+\delta_\text x^2},\\
  \langle\hat A_{22}\rangle&=\frac{z_\text x+\tfrac{P_\text x}{\Gamma+P_\text x}}{1+2z_\text x},\\
  |\langle\hat A_{12}\rangle|^2&=\frac{z_\text x}{(1+2z_\text x)^2}\left(\frac{\Gamma-P_\text x}{\Gamma+P_\text x}\right)^2.
\end{align}
Similarly to dephasing, the excitation is increased while the coherence is decreased by the incoherent pumping. 
Restricting
$P_\text x\leq\Gamma$, the saturation case is $P_\text x=\Gamma$, for which $\langle\hat A_{22}\rangle=1/2$ and $|\langle\hat A_{12}\rangle|^2=0$, independent of the coherent pumping from the laser. We emphasize, that, when defining the quantity $P=P_\text x/(\Gamma+P_\text x)>0$, we can write the solutions as
\begin{align}
  \langle\hat A_{22}\rangle&=\frac{z_\text x+P}{1+2z_\text x},\\
  |\langle\hat A_{12}\rangle|^2&=\frac{z_\text x(1-2P)^2}{(1+2z_\text x)^2}.
\end{align}
The term $P$ appears in place of the purification rate $R$, compare Eqs.~(\ref{eq.ana-A22sol}), (\ref{eq.ana-A12sol}), but with opposite sign, so that the purity of the quantum state of the SPE decreases. 

In the cavity-assisted scenario one may expect $R$ and $P$ to directly counteract each other. Complementing the calculations from Eqs.~(\ref{eq.ana-a})-(\ref{eq.ana-A12sol}) by incoherent pumping yields
\begin{align}
  V_\text x&=i\delta_\text x+\frac{\Gamma+P_\text x}{2}+\frac{g^2}{i\delta_\text c+\tfrac{\kappa}{2}},\label{eq.Vx}\\
  z_\text x&=\frac{2\Omega_\text R^2}{\Gamma+P_\text x}\frac{\Re[V_\text x]}{|V_\text x|^2}\approx\frac{\Omega_\text R^2}{|V_\text x|^2}\approx z,\label{eq.zx}\\
  R_\text x&=\frac{\kappa}{\Gamma+P_\text x}\langle\hat a^\dagger\hat a\rangle,\label{eq.Rx}\\
  \langle\hat A_{22}\rangle&=\frac{z_\text x+P-R_\text x}{1+2z_\text x},\label{eq.ana-A22xsol}\\
  |\langle\hat A_{12}\rangle|^2&=\frac{z_\text x(1-2P+2R_\text x)^2}{(1+2z_\text x)^2}\label{eq.ana-A12xsol}.
\end{align}
The excitation parameter $z_\text x\approx z$ does not change significantly. Likewise the structure of the expectation values itself is identical to the case of no incoherent pumping, if we define
\begin{equation}
  \tilde R_\text x=R_\text x-P=\frac{\kappa\langle\hat a^\dagger\hat a\rangle-P_\text x}{\Gamma+P_\text x}
\end{equation}
as the new purification rate of the cavity setup. For $\tilde R_\text x>0$ we have purification, $\tilde R_\text x=0$ corresponds to the free-space scenario, 
and for $\tilde R_\text x<0$ we impurify the state. 
Consequently, squeezing behaves as in the case of no incoherent pumping, Eq.~(\ref{eq.Sq-ana}), but with $\tilde R_\text x$ replacing $R$.

Two consequences of the solutions~(\ref{eq.Vx})-(\ref{eq.zx}) should be noted. First, when the SPE is tuned through a cavity resonance the sign of $\tilde R_\text x$ changes from negative to positive and back, changing the behavior decreasing to increasing purity of the atomic state and back. Compared with the simple cavity-assisted case, where we have purification, as $R>0$, here we can  control the purity of the quantum state of the SPE by tuning its resonance frequency. 
Second, the effect of incoherent pumping is limited by $P_\text x\leq\Gamma$ or equivalently $P\leq1/2$. The saturation case $P=1/2$ is, however, only a theoretical value. For example in case of phonon-induced pumping, this is equal to infinite temperature. For our scenario with $\kappa\gg\Gamma>P_\text x$, the purification and thus optimized squeezing is nearly unaffected by the incoherent pumping, at least at the cavity resonance. 

\begin{figure}[h]
  \includegraphics[width=8cm]{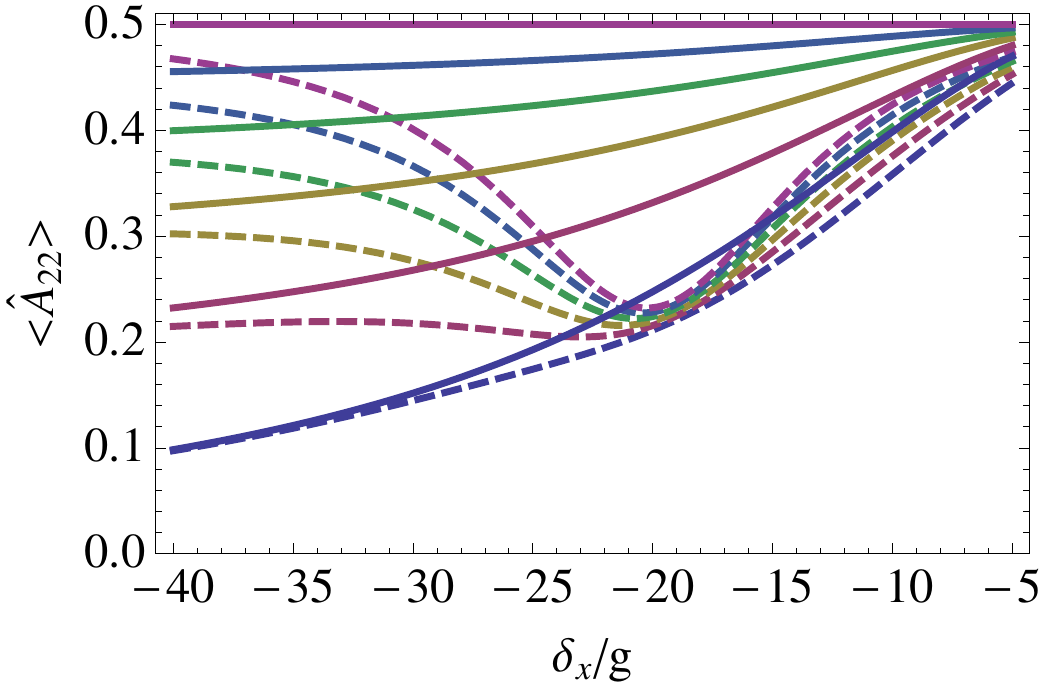}
  \caption{(color online). As Fig.~\ref{fig.A22Deph}, for different incoherent pumpings $P_\text x$. The solid lines represent the free space case, the dashed ones the corresponding cavity assisted scenario. From bottom to top (for each solid and dashed lines separately): $P_\text x/\Gamma=0,0.2,0.4,0.6,0.8,1$. All other parameters are as in Fig.~\ref{fig.Sqmax}.}\label{fig.A22incx}
\end{figure}

\begin{figure}[h]
  \includegraphics[width=8cm]{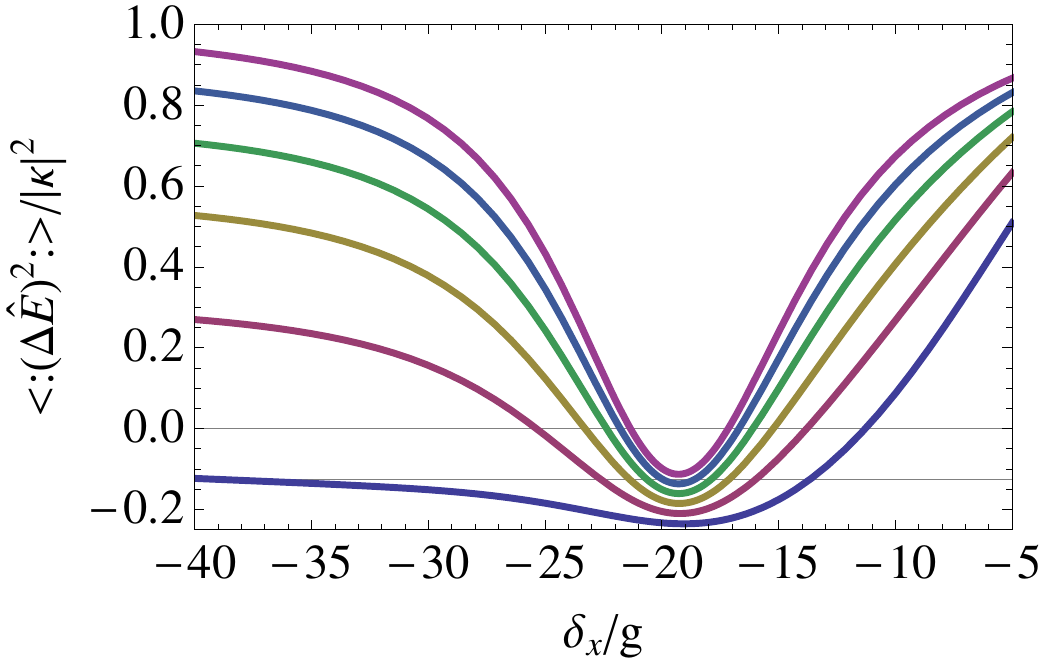}
  \caption{(color online). Squeezing of the SPE fluorescence over $\delta_\text x$ for different incoherent pumpings $P_\text x$. From bottom to top: $P_\text x/\Gamma=0,0.2,0.4,0.6,0.8,1$. All other parameters are as in Fig.~\ref{fig.Sqmax}. The horizontal lines indicate maximal free space squeezing (-1/8) and vanishing squeezing (0).}\label{fig.Xpump}
\end{figure}

In Fig.~\ref{fig.A22incx} we compare $\langle\hat A_{22}\rangle$ for different values of $P_\text x$ inside and outside the cavity. Similar to the case of dephasing, the pronounced cavity resonance effect indicates an increase of $\langle\hat a^\dagger\hat a\rangle$, cf. Eqs.~(\ref{eq.Rx}), (\ref{eq.ana-A22xsol}). Remarkably, even for the saturated scenario, in the cavity resonance, the excitation of the SPE remains significantly below the free space value for no incoherent pumping.

In Fig.~\ref{fig.Xpump} we show the squeezing for different incoherent pumping rates $P_\text x$. 
The squeezing at the cavity resonance is quite robust against incoherent pumping, while off resonance it is quickly lost. 
Even for saturated pumping, $P_\text x=\Gamma$, we obtain significant squeezing. At $\delta_\text x=-19.3 g$, the phase-optimized normally ordered variance, Eq.~(\ref{eq.Atvarsol1}), attains a value of $-0.113$, which is close to the  maximal possible squeezing of $-1/8$ in free-space fluorescence. For saturated incoherent pumping in free space, the normally ordered variance would be at $+1$ for all parameters, so that squeezing is impossible.

\subsection{Incoherent Pumping of the Cavity}
The cavity mode may also be incoherently pumped, either directly from the phonons or from the interaction with the SPE.  Nevertheless, this effect is expected to be smaller than the interaction of the  cavity mode with phonons. The latter was observed to be negligibly small~\cite{Ahrens}. 
The cavity is supposed to be pumped incoherently with a rate $P_\text c\leq\kappa$, where equality again represents saturation. Since the cavity mode is bosonic, this implies $\langle\hat a^\dagger\hat a\rangle\rightarrow\infty$. As we are interested in the case $\langle\hat a^\dagger\hat a\rangle\ll1$, we are limited to $P_\text c\ll\kappa$.

After applying the formalism of Eqs.~(\ref{eq.ana-a})-(\ref{eq.ana-A12sol}) for this scenario, we obtain terms, which resemble the previous case of incoherent SPE pumping:
\begin{align}
  V_\text c&=i\delta_\text x+\frac{\Gamma}{2}+\frac{g^2}{i\delta_\text c+\tfrac{\kappa-P_\text c}{2}},\label{eq.Vc}\\
  z_\text c&=\frac{2\Omega_\text R^2}{\Gamma}\frac{\Re[V_\text c]}{|V_\text c|^2}\approx\frac{\Omega_\text R^2}{|V_\text c|^2}\approx z,\label{eq.zc}\\
  R_\text c&=\frac{\kappa-P_\text c}{\Gamma}\langle\hat a^\dagger\hat a\rangle,\label{eq.Rc}\\
  \langle\hat A_{22}\rangle&=\frac{z_\text c+P_\text c/\Gamma-R_\text c}{1+2z_\text c},\label{eq.ana-A22csol}\\
  |\langle\hat A_{12}\rangle|^2&=\frac{z_\text c(1-2P_\text c/\Gamma+2R_\text c)^2}{(1+2z_\text c)^2}.\label{eq.ana-A12csol}
\end{align}
Again, we can define an effective purification rate
\begin{equation}
	\tilde R_\text c=R_\text c-\frac{P_\text c}{\Gamma}=\frac{(\kappa-P_\text c)\langle\hat a^\dagger\hat a\rangle-P_\text c}{\Gamma},
\end{equation}
which is formally similar to $\tilde R_\text x$. However, contrary to $P_\text x$, $P_\text c$ is not limited by $\Gamma$ but only by $\kappa$ which is very large compared to $\Gamma$. Hence, we may have $P_\text c>\Gamma$, without violating $\kappa\gg P_\text c$. In such a case however, the incoherent pumping contributes strongly to $\tilde R_\text c$ and the squeezing in the cavity resonance is suppressed. On the other hand, out off the cavity resonance the effective coupling between SPE and cavity mode is very weak, which yields $\tilde R_\text c\approx0$. As the environmental effect is only caused by the cavity, we obtain the free space scenario again with the remaining squeezing of weak effective pumping $\delta_\text x^2\gg\Omega_\text R^2$. Of course, in this case the atomic-state purification does not occur. 

\begin{figure}[h]
  \includegraphics[width=8cm]{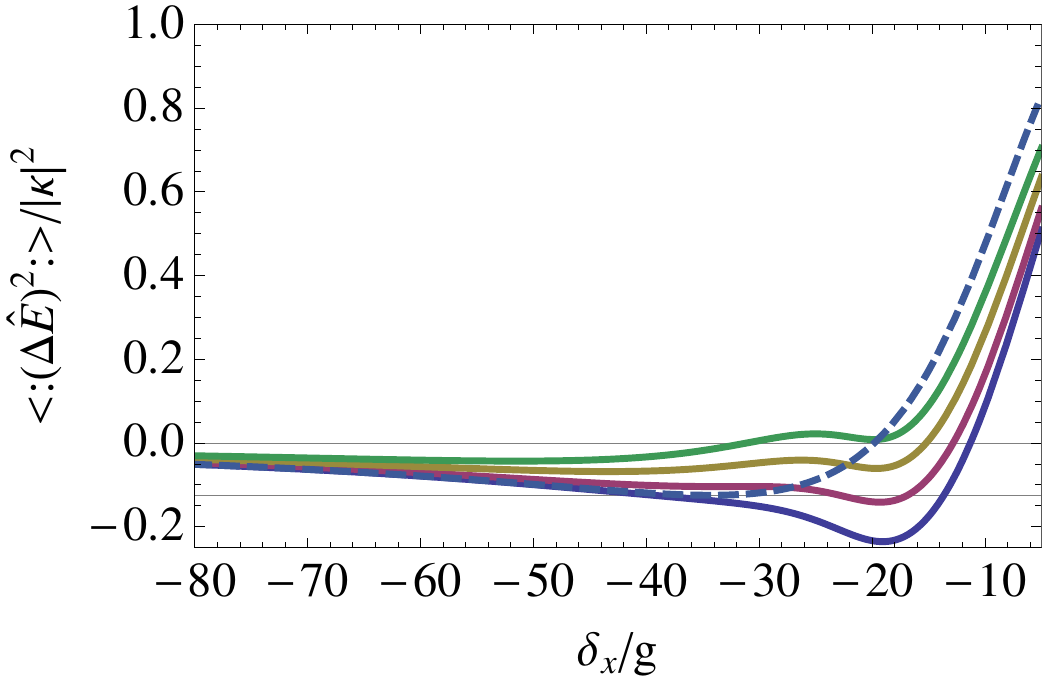}
  \caption{(color online). Squeezing of the SPE fluorescence over $\delta_\text x$ for different incoherent pumpings $P_\text c$. From bottom to top (solid lines): $P_\text c/\Gamma=0,1,2,3$. The dashed line is the free space fluorescence squeezing ($g=0$). All other parameters are as in Fig.~\ref{fig.Sqmax}. The horizontal lines indicate maximal free space squeezing (-1/8) and vanishing squeezing (0).}\label{fig.Cpump}
\end{figure}

In Fig.~\ref{fig.Cpump}, we show the squeezing (note the larger region of $\delta_\text x$) for different incoherent pumping rates. For comparison, the free-space squeezing is also given. For large $|\delta_\text x|$-values we approach the free space value, while in the cavity resonance the squeezing is suppressed significantly for increasing $P_\text c$. However, for an incoherent pumping equal to the spontaneous emission of the SPE, we still obtain squeezing of about the maximal free-space value.

\section{Detection of Squeezing}\label{sec.VI}
The prediction of squeezing in the resonance fluorescence of a SPE could not be confirmed in experiments yet.
Usually the normally ordered variance of a light field is measured by balanced homodyne detection, for details see e.g.~\cite{WelVo}. In the case of single-atom fluorescence the following problems must occur. First, the atomic motion yields phase shifts, which can be eliminated  by using trapped ions or well localized excitations in semiconductor systems. 
Second, the small collection efficiency of the field substantially reduces the observable squeezing. 
This problem can be resolved by homodyne correlation measurements~\cite{vogel,vogel95,shchukin}.

\begin{figure}[h]
  \includegraphics[width=6cm]{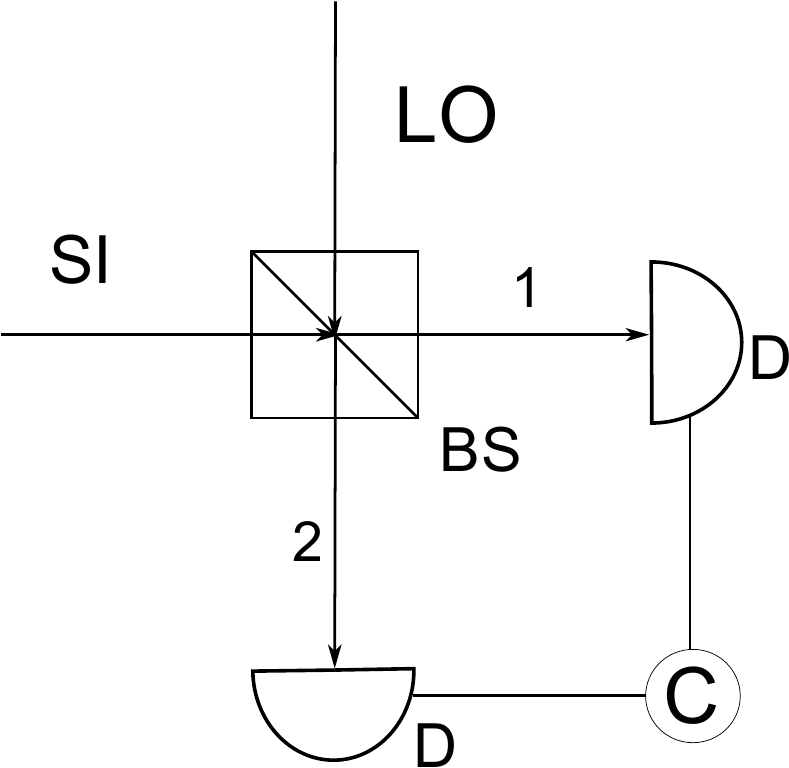}
  \caption{Setup for the homodyne cross-correlation measurement of the
signal field $\text{SI}$, $\text{LO}$ denotes the local oscillator. The fields are combined by a beamsplitter (BS) and the superimposed fields 1 and 2 are measured by correlating C the two detectors D; after~\cite{vogel95}.}\label{fig.homcross}
\end{figure}

Let us reconsider the homodyne cross-correlation measurement~\cite{vogel95}, the setup is shown in Fig.~\ref{fig.homcross}. The signal field $\hat E_\text{SI}$ is superimposed by a beam splitter with the coherent local oscillator field of amplitude $E_\text{LO}$. The cross-correlation between the two outgoing light fields $\hat E_1$ and $\hat E_2$ is recorded. The measured signal,
\begin{equation}
  \mathcal G^{(2,2)}(t_1,t_2)=\eta^2\langle\hat E_1^{(-)}(t_1)\hat E_2^{(-)}(t_2)\hat E_2^{(+)}(t_2)\hat E_1^{(+)}(t_1)\rangle,
\end{equation}
is the given by the intensity cross-correlation function and the (equal) quantum efficiencies $\eta$ of the two detectors. Note that $\eta$ includes the (small) collection efficiency of the fluorescence signal.
Following~\cite{vogel95}, for equal times $(t_1=t_2=t)$ we get
\begin{equation}
\begin{split}
  \mathcal G^{(2,2)}(t,t)=\frac{\eta^2}{4}[\langle:\hat I_\text{SI}^2:\rangle+I_\text{LO}^2-2I_\text{LO}\Re(\langle\hat E_\text{SI}^{(+)2}\rangle)],\label{eq.G22-sum}
\end{split}
\end{equation}
with $\hat I_\text{SI}=\hat E^{(-)}_\text{SI}\hat E^{(+)}_\text{SI}$ and $I_\text{LO}=E_\text{LO}^2$. For a sufficiently large time delay 
$(t_2-t_1\rightarrow\infty)$ we approach the uncorrelated events,
\begin{equation}
\begin{split}
  \mathcal G^{(2,2)}_\text{unc}(t)=&\frac{\eta^2}{4}[\langle\hat I_\text{SI}\rangle^2+I_\text{LO}^2\\
    &-2I_\text{LO}\left(\Re(\langle\hat E_\text{SI}^{(+)2}\rangle)+|\langle\hat E_\text{SI}^{(+)}\rangle|^2-\langle\hat I_\text{SI}\rangle\right)].
\end{split}
\end{equation}

Let us consider the difference $\Delta{\mathcal G}^{(2,2)}$ of ${\mathcal G}^{(2,2)}(t,t)$ and ${\mathcal G}^{(2,2)}_\text{unc}(t)$,  
\begin{equation}
  \Delta\mathcal G^{(2,2)}=\frac{\eta^2}{4}\left(\langle: (\Delta \hat I_\text{SI})^2:\rangle
-I_{\text{LO}}\langle:(\Delta\hat E_{\text{SI}})^2:\rangle\right).
\end{equation}
This difference includes the normally ordered variances of the intensity and the field strength of the signal. 
For a SPE we have 
$\langle:\hat I_\text{SI}^2:\rangle=0$, so that we obtain
\begin{equation}
  \Delta\mathcal G^{(2,2)}=-\frac{\eta^2}{4}\left(I_{\text{SI}}^2+I_{\text{LO}}\langle:(\Delta\hat E_{\text{SI}})^2:\rangle\right).\label{eq.DelGfin}
\end{equation}
Only if the field is squeezed, $\langle:(\Delta\hat E_{\text{SI}})^2:\rangle<0$, $\Delta\mathcal G^{(2,2)}$ can become positive.  For squeezed fields and sufficiently strong local oscillator, $\Delta\mathcal G^{(2,2)}$ switches the sign for some phase of the signal, and squeezing is detected by $\Delta\mathcal G^{(2,2)}>0$. 

From Eq.~(\ref{eq.DelGfin}) it may seem that a strong local oscillator is preferential for detection the squeezing. However, since the present simple detection scheme is not balanced, the classical fluctuations of the local oscillator must be considered~\cite{OHM}. In fact, this problem can be avoided by using a more complex balanced homodyne correlation setup~\cite{shchukin}. For the present scheme,
the dominant classical noise term is 
\begin{equation}
  \Delta\mathcal G_{\rm cl}^{(2,2)}=\eta^2 I_{\text{LO}} \overline{(\delta  E_{\text{LO}})^2},\label{eq.class}
\end{equation}
where $\overline{(\delta  E_{\text{LO}})^2}$ is the classical amplitude variance of the local oscillator. This classical noise
is easily measured by blocking the signal channel. The squeezing condition finally becomes
\begin{equation}
  \Delta\mathcal G^{(2,2)}>\Delta\mathcal G_{\rm cl}^{(2,2)}.\label{eq.sq-class}
\end{equation}
Following the discussion in~\cite{vogel95}, the optimal experiment is performed with a weak local oscillator, whose intensity slightly exceeds the intensity of the fluorescence signal of the SPE.

\section{Conclusions and Outlook}\label{sec.VII}
We have studied the optimization of squeezing in resonance fluorescence of a single-photon emitter through cavity-assisted purification of the atomic quantum state. This can be achieved by   tuning the cavity on resonance with the lower Rabi-sideband of the emitter. The squeezed light is recorded out the side of the cavity.
The maximal squeezing is significantly larger and more robust against 
disturbance than in free space. 

Analytical approximations are given, which yield an interpretation of the basic effects of our purification scenario.
A detailed study is given of the resistance of the optimized squeezing against environmental disturbances. In particular, it is shown that squeezing is much more robust against dephasing and incoherent pumping compared with an atom in free space. Consequently, even strong incoherent channels do not fully suppress the squeezing in our optimized setting. All the considered incoherent effects are present for quantum dots in semiconductor microcavities. Our results indicate that such complex devices may be promising integrated squeezed-light sources. A simple homodyne correlation measurement technique is considered, which renders it possible to detect the squeezing of a laser-driven single-photon emitter.

It is of some interest to compare the relation of the squeezed light sources under study with standard sources based on optical parametric 
oscillators. Both types of sources are very different from two perspectives. The squeezing in resonance fluorescence is diminished by the small collection efficiency, which is not the case for standard sources. On the other hand, our observation scheme is not sensitive to the efficiency, whereas the balanced homodyne detection in the standard case is. Hence, it is a challenging open problem to compare the advantages of both scenarios for practical applications. However, this problem is beyond the scope of our paper as it requires further research.
 
\acknowledgments This work was supported by the Deutsche Forschungsgemeinschaft through SFB~652.

\appendix
\section{Numerical calculations.}\label{sec.appnum}
The master equation for a SPE in a single mode cavity yield an infinite hierarchy of coupled equations for the density matrix elements
\begin{equation}
  \varrho_{n,i;m,j}=\langle n,i|\hat \varrho|m,j\rangle.
\end{equation}
Here, the first index is the cavity photon number and the second is the SPE excitation number ($i,j=1,2$). The explicit equations for the cavity assisted system, without further environmental effects, can be written as
\begin{widetext}
\begin{align}
 \dot\varrho_{n,1;m,1}=&-[i\delta_\text c(n-m)+\tfrac{\kappa}{2}(n+m)]\varrho_{n,1;m,1}-ig[\sqrt{n}\varrho_{n-1,2;m,1}-\sqrt{m}\varrho_{n,1;m-1,2}]\nonumber\\
		     &-i\Omega_\text R[\varrho_{n,2;m,1}-\varrho_{n,1;m,2}]+\Gamma\varrho_{n,2;m,2}+\kappa\sqrt{(n+1)(m+1)}\varrho_{n+1,1;m+1,1},\\
\dot\varrho_{n,1;m,2}=&[i(\delta_\text a-(n-m)\delta_\text c)-\tfrac{\Gamma+\kappa(n+m)}{2}]\varrho_{n,1;m,2}-ig[\sqrt{n}\varrho_{n-1,2;m,2} -\sqrt{m+1}\varrho_{n,1;m+1,1}]\nonumber\\
		      &- i\Omega_\text R(\varrho_{n,2;m,2}-\varrho_{n,1;m,1})+\kappa\sqrt{(n+1)(m+1)}\varrho_{n+1,1;m+1,2},\\
  \dot\varrho_{n,2;m,1}=&-[i(\delta_\text a+(n-m)\delta_\text c)+\tfrac{\Gamma+\kappa(n+m)}{2}]\varrho_{n,2;m,1}-ig[\sqrt{n+1}\varrho_{n+1,1;m,1} -\sqrt{m}\varrho_{n,2;m-1,2}]\nonumber\\
		      &- i\Omega_\text R(\varrho_{n,1;m,1}-\varrho_{n,2;m,2})+\kappa\sqrt{(n+1)(m+1)}\varrho_{n+1,2;m+1,1},\\
  \dot\varrho_{n,2;m,2}=&-[i\delta_\text c(n-m)+\Gamma+\tfrac{\kappa}{2}(n+m)]\varrho_{n,2;m,2}-ig[\sqrt{n+1}\varrho_{n+1,1;m,2}-\sqrt{m+1}\varrho_{n,2;m+1,1}]\nonumber\\
		      &-i\Omega_\text R(\varrho_{n,1;m,2}-\varrho_{n,2;m,1})+\kappa\sqrt{(n+1)(m+1)}\varrho_{n+1,2;m+1,2}.
\end{align}
\end{widetext}
We truncate the set of equations at a sufficiently large photon number $N$. By varying $N$, the validity of the calculations can be checked. Using $\text{Tr}\{\hat \varrho\} =1$, we can eliminate one element of the main diagonal, in our case, we choose $\varrho_{0,1;0,1}$.  This introduces an inhomogeneity into the equations, allowing us to calculate the steady state density matrix simply by  inverting the matrix of coefficients and multiplying with the inhomogeneity. Finally, the expectation values of interest can be directly obtained.

\end{document}